\documentclass[aps,prb,floatfix,twocolumn,10pt]{revtex4-1}
\usepackage[pdftex]{graphicx}
\usepackage{amsmath,amssymb,subfigure,color,verbatim} 

\providecommand{\myheading}[1]{\textbf{#1}}
\providecommand{\lesssim}{asdfasdf}
\providecommand{\text}[1]   {\mathrm{#1}}
\providecommand{\tfrac}[2]  {\textstyle{\frac{#1}{#2}}}

\providecommand{\MMM}{\mathbf{M}}

\providecommand{\QQQ}{\mathbf{Q}}
\providecommand{\RRR}{\mathbf{R}}
\providecommand{\SSS}{\mathbf{S}}

\providecommand{\calL}{\mathcal{L}}

\providecommand{\dd}{\partial}
\providecommand{\percent}{\%{}}

\providecommand{\up}{\uparrow}
\providecommand{\dn}{\downarrow}

\providecommand{\mean}[1]{\left\langle #1 \right\rangle}

\providecommand{\ket}[1]{\left|#1\right>}
\providecommand{\bra}[1]{\left<#1\right|}
\providecommand{\braket}[2]{\left< #1 \right. \left| #2 \right>}

\begin{document}
\title{Cooling by corralling: a route to ultra-low entropies in optical lattices}
\author{Yen Lee Loh}
\affiliation{Department of Physics and Astrophysics, University of North Dakota, Grand Forks, ND  58202, USA}
\begin{abstract}
A major motivation for cold atom experiments
	is the search for quantum ground states
	such as antiferromagnets and d-wave superfluids.
The primary obstacle to this task 
	is the difficulty of cooling to sufficiently low temperatures.
We propose a way to achieve very low temperatures and entropies
($\sim 0.03k_B$ per particle)
	by trapping fermions in a corral formed from another species of atoms.
The Fermi system can then be used as a heat sink,
	or it can be adiabatically evolved into other desired states.
In particular, we suggest methods for generating antiferromagnetism 
	using this technique.
\end{abstract}
\date{Dec 13, 2011}
\maketitle


The last decade has seen astonishing progress in ultracold atom technology.  
In particular, atomic gases in optical lattices can now be imaged and manipulated with amazing controllability and fidelity \cite{bakr2009,bakr2010,sherson2010,weitenberg2011},
in a way that parallels the development of atomic-resolution scanning tunneling microscopy in the 1980s.
Recent advances in the Harvard and Munich groups \cite{bakr2010,weitenberg2011,bakr2011oeb} have succeeded in creating high-quality Bose Mott insulators (MIs).  The core of such a MI contains one boson per lattice site; there are few defects, and thus the entropy density is very low.

This is a promising starting point from which to apply an adiabatic protocol to access an interesting quantum ground state, but there are two caveats. 
First, the initial state must not only have low entropy, but it must be close to the \emph{equilibrium} ground state of the initial Hamiltonian.  Low-entropy non-equilibrium states created using orbital excitation blockade (OEB) techniques \cite{bakr2011oeb} or single-site addressing \cite{weitenberg2011} do not satisfy this requirement on their own.
Second, the experiments mentioned above have the MI core surrounded by a superfluid (SF) shell.  When the lattice depth is reduced during the adiabatic protocol, entropy from the SF will leak inwards and contaminate the MI.  To prevent this, the MI must be \emph{isolated}.

Cooling by trap shaping \cite{popp2006,bernier2009,williams2010,mathyArxiv2012} is an attractive way to overcome both these obstacles.
In this approach one typically uses a deep ``dimple'' to confine a gapped phase with a low ``entropy capacity'', surrounded by a shallow ``entropy storage region'' containing a gapless phase of high ``entropy capacity'', such that at thermal equilibrium the bulk of the entropy resides in the storage region.  A barrier potential is used to isolate the low-entropy core from the high-entropy shell.  The core region is then an isolated quantum system close to the ground state of its Hamiltonian, as desired.


An important limitation on this approach is how precisely one can control the barrier potential.  Typical proposals use optical methods, e.g., an annular laser beam \cite{bernier2009}.  Recent advances in laser beam shaping using digital micromirror devices have been able to create flattop beam profiles with less than $1\percent$ RMS error \cite{liangProc2011}, corresponding to square potential wells with relatively sharp walls.  
However, optical beam shaping is inherently diffraction-limited and inevitably produces a boundary region a few sites thick, which contains unwanted entropy \cite{bernier2009}.

	\begin{figure}[!htb]   \centering
		\includegraphics[width=0.45\textwidth]{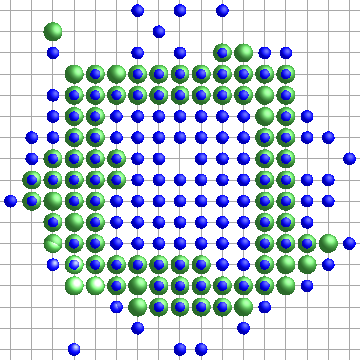}
		\caption{
		\label{FermionsInCorral}
			\textbf{Fermions in a bosonic corral:}
			Large green spheres represent bosons and
			small blue spheres represent fermions.
			The grid represents the optical lattice.
			The fermions are trapped in the corral and are unable to tunnel out.
			In the absence of vacancies (such as the one purposely included in this illustration)
				the fermions form a zero-entropy, zero-temperature ground state
				of an isolated system.
			This is like using a cookie cutter
				to isolate a clean, homogeneous piece of a slab of dough.
		}
	\end{figure}
	
\myheading{Preparation of low-entropy states:}
Here we propose a novel method for creating isolated atomic gases in low-entropy equilibrium states.
The key idea is to form a corral using \emph{atoms} instead of light.  The resulting potential well has atomically sharp edges, and so the boundary region is essentially eliminated.

\begin{figure}[!htb] \centering
		\includegraphics[width=0.7\columnwidth]{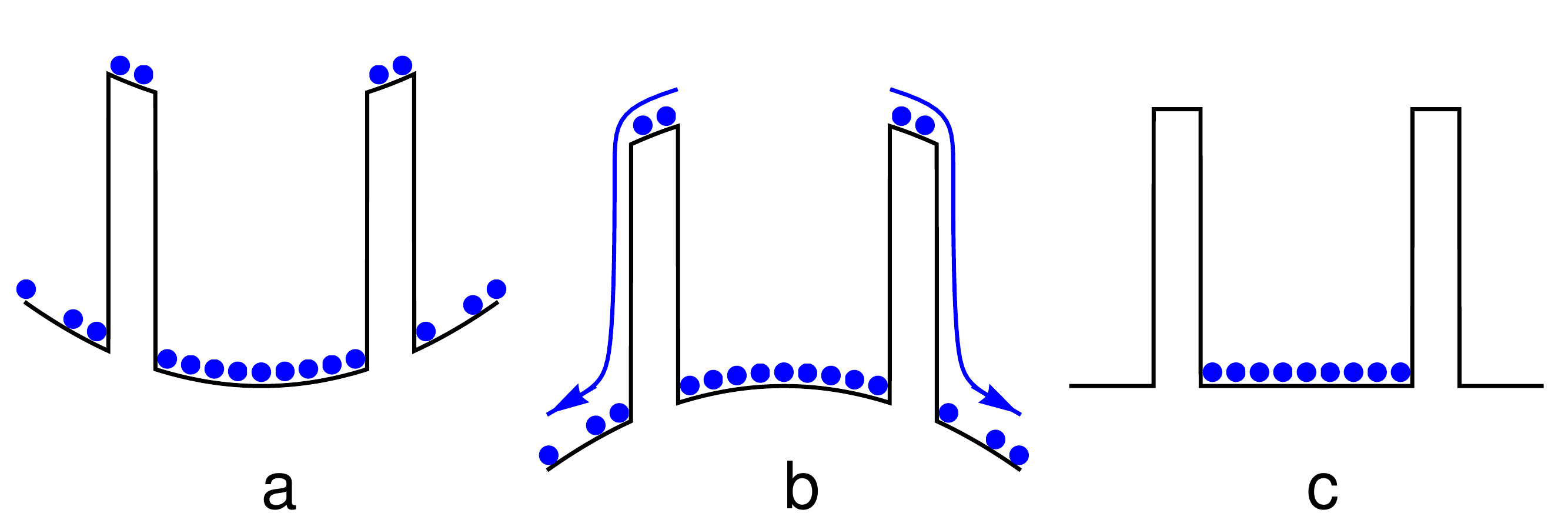}
		\caption{
		\label{TrapInversion}
			\textbf{Discarding excess fermions by trap inversion:}
			(a) First, the Fermi insulator is formed in a harmonic trap,
				the bosonic corral is created,
				and the boson-fermion interaction is ramped up.
				Black curves show the combination of the trap potential, $V^\text{trap}_f$,
					and the Hartree potential, $U_\text{bf} n_b$, seen by the fermions.
			(b) The trap potential is now ``turned upside down''
				to ``pour away'' unwanted fermions.
			(c) Finally, the inverted trap potential is turned off 
				to give a flat-bottomed infinite square well.
			The lattice and $z$-confinement are maintained 	throughout this process.
			The process is akin to throwing away unused dough
				after cutting out a cookie of the desired shape.
		}
	\end{figure}

\begin{figure}[h]  \centering
		\includegraphics[width=0.9\columnwidth]{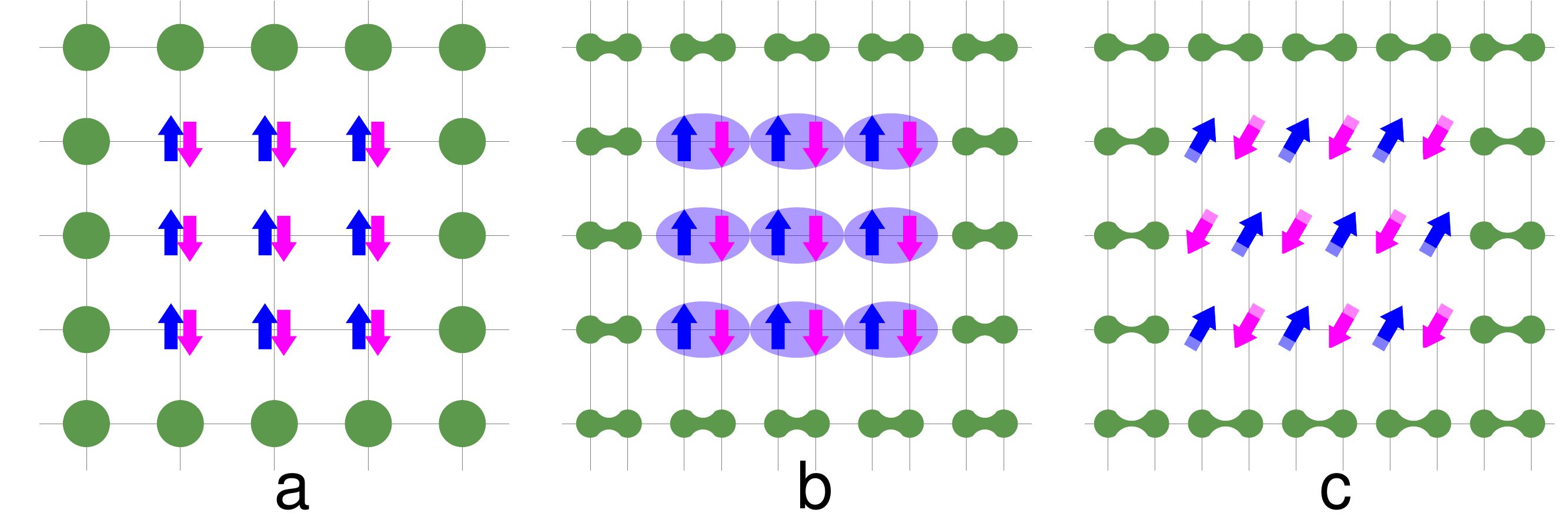}
		\caption{
		\label{SiteDoubling}
			\textbf{Site-doubling route to antiferromagnetism:}
			A square lattice is loaded with two fermions per site,
				forming a band insulator (a).
			Each site of the lattice is then split into a double well.
			The ground state consists of singlets in each double well (b).
			As the hopping between double wells ($t'$) increases,
				antiferromagnetic correlations grow,
				and at a certain value of $t'/t$ there is a quantum phase transition
				to AF order.
		}
	\end{figure}

This idea may be implemented in 1D, 2D, or 3D; we will focus on the 2D case.
A single species of fermions is loaded into a trap,
and an optical lattice is gradually ramped up.  
(The trap is deep enough to produce a band insulator (BI) core within a Fermi liquid (FL) shell, and the lattice is deep enough that only the lowest Bloch band is occupied.)
A bosonic Mott insulator 
	is loaded into the same spatial lattice 
	and patterned into a hollow 2D shape \cite{bakr2010,sherson2010,weitenberg2011}.
Experimental parameters are chosen such that the bosons experience a deep lattice, so that boson tunneling is negligible
and the bosons stay in place for the remainder of the experiment~\footnote{
The atom species, laser parameters, and $B$ field should be chosen such that the fermions experience a relatively shallow lattice ($V^\text{latt}_f/E^R_f \lesssim 8$) while the boson lattice depth is much higher ($V^\text{latt}_b/E^R_b \gtrsim 20$). 
Then the fermion hopping amplitude is appreciable, allowing the processes in Figs.~\ref{TrapInversion} and \ref{SiteDoubling} to occur on a realistic timescale, whereas the boson hopping is exponentially slower and is negligible over the duration of the experiment.
}.
The boson-fermion interaction $U_\text{bf}$ is now ramped up to a large repulsive value.
The bosons form a potential barrier of height $U_\text{bf}$ that is impermeable to the fermions.
The $N$ fermions within this ``corral'' now constitute an isolated quantum system with a density close to 1 fermion per site and a very low entropy (see Fig.~\ref{FermionsInCorral}).


\myheading{Corral shape and preparation:}
The corral of bosons may be prepared using the method of Ref.~\onlinecite{weitenberg2011}, where a tightly focused laser beam, together with a microwave field, was used to flip the hyperfine spin of individual atoms in a singly-occupied bosonic Mott insulator with sub-diffraction-limited resolution well below the lattice spacing, and a push-out laser pulse was then applied to remove the flipped (or unflipped) atoms.

Alternatively, as in Refs.~\onlinecite{bakr2010,sherson2010}, one may begin with a large bosonic system consisting of a doubly-occupied Mott insulating core surrounded by a singly-occupied Mott insulating ring, and then illuminate the atoms to eject pairs of atoms via light-assisted collisions (photoassociation).  This results in a core of empty sites surrounded by a MI ring, which can then be used as the corral.
Although this technique provides less control over the exact shape of the corral, it has the advantage that it may be used to produce a \emph{spherical} corral which can then be used to make a 3D antiferromagnet (AF), which is much more robust than its 2D analog.
Yet another possibility is to prepare a boson-fermion mixture with trapping potentials and interactions such that the fermion cloud occupies the center of the trap and the bosons are expelled to form an outer shell -- i.e., the fermion cloud itself is used to form a ``mold'' which is then used to ``cast'' it.
The latter approaches have the advantage of not requiring single-site resolution.
\footnote{
There are many other possibilities; for example, the method may work equally well with \emph{attractive} boson-fermion interactions.
}.

An appealing feature of our proposal is that the corral does not have to be in the shape of a perfect square, circle, or sphere.  The only requirement is that the bosons form a multiply connected region with a continuous wall enclosing some empty sites.  The digital (in this case, binary) nature of the boson occupation number leads us to a zero-entropy fermionic state, regardless of irregularities, as illustrated in Fig.~\ref{FermionsInCorral}.

In order to confine the Fermi band insulator, the corral potential height $U_\text{bf}$ should be chosen to be greater than the bandwidth.
Also, to prevent fermion loss by quantum tunneling through the walls, and to minimize the probability of gaps and leakage, the corral should be made several bosons thick.

\myheading{Discarding excess fermions:}
Although the interior of the corral is already an isolated system, 
it is nevertheless a good idea to get rid of the excess fermions on and outside the corral
to prevent them from interfering with later stages of the experiment (see discussion in Ref.~\onlinecite{bernier2009}).  We first present a simple picture of how this could be done.
Figure~\ref{TrapInversion}a shows a slice of Fig.~\ref{FermionsInCorral} in the $x$-direction.
Up to this point the fermions are still under harmonic confinement.
Now, the fermionic trap potential in the $(x,y)$ plane is \emph{inverted} (while maintaining $z$-confinement).
Then, the exterior fermions accelerate outward and rapidly exit the system, whereas the interior fermions remain confined by a potential barrier that is too thick for them to tunnel through.  
(It may be advantageous to lower the optical lattice depth to facilitate this separation.)
After a suitable time, the effective trapping frequencies (including contributions from any compensating lasers, field gradients, etc.) are reset to zero in the $x$ and $y$ directions.
This leads to the ideal situation of fermions in a flat-bottomed square well with infinitely high walls.

Obviously, optimizing the protocol is a complicated task that depends on precise details of the system being considered; this is a topic of further work.  Nevertheless, preliminary simulations indicate that the idea is realistic and feasible (see supplementary information and ancillary file containing animation).

\myheading{Adiabatic evolution to interesting ground states:}
A low-entropy, low-temperature system can be used as a heat sink to cool another system.
However, the single-species Fermi insulator described above would be an inefficient heat sink (due to its excitation gap, low heat capacity, low thermal conductivity, and limited phase space); BECs and other fluids generally function much better as heat sinks.  It is therefore most desirable to look for ways to adiabatically evolve the Fermi BI to an interesting ground state.  In the remainder of this paper we present a way to attain an antiferromagnetic state.

	\begin{figure}[h]  \centering
		\includegraphics[width=0.7\columnwidth]{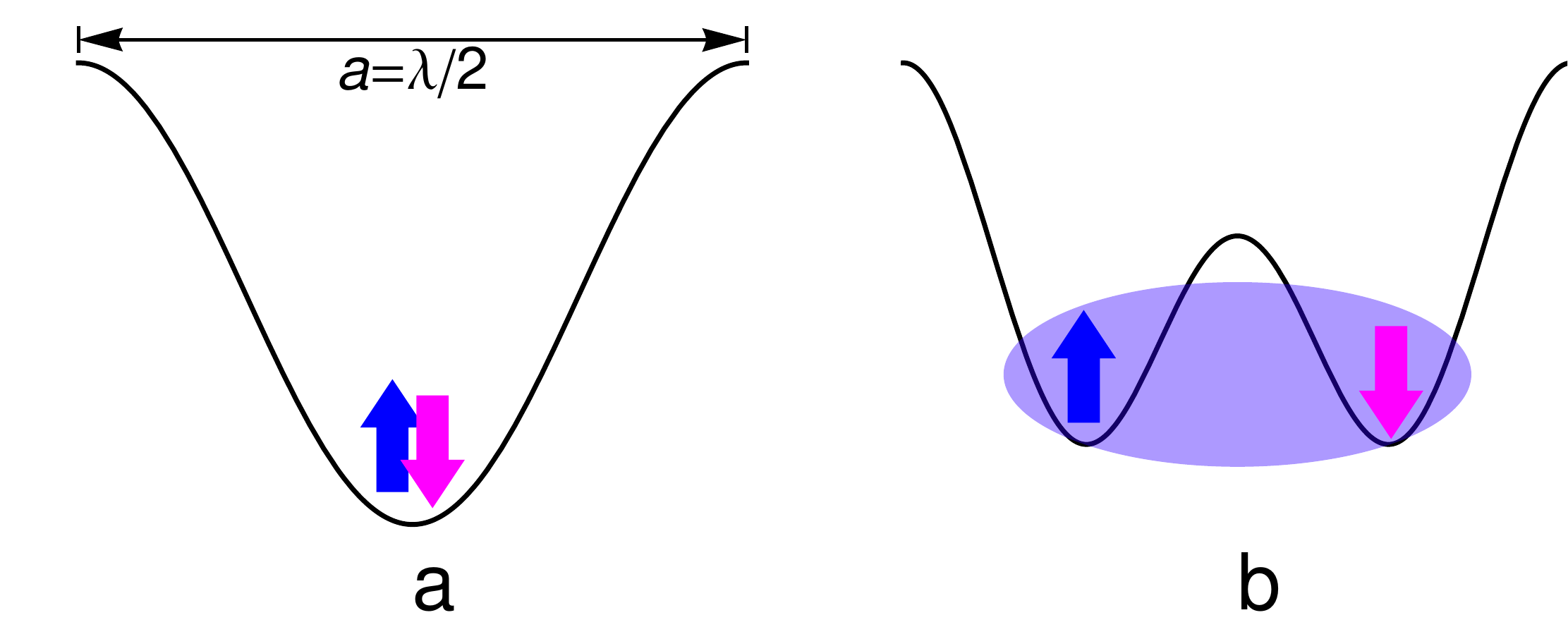}
		\caption{
		\label{DoubleWell}
			\textbf{Splitting a singlet using a double-well potential:}
			Beginning with a pair of fermions in a well of a deep optical lattice,
				which is then adiabatically morphed into a double-well shape,
				one obtains a two-site singlet 
				$\frac{1}{\sqrt{2}} 
				\big(  \ket{\up}_L \ket{\dn}_R - \ket{\dn}_L \ket{\up}_R \big)$.
		}
	\end{figure}

\myheading{Protocol for obtaining a 2D antiferromagnet:}
\emph{Two} species of fermions ($\up$ and $\dn$) are loaded into a bosonic corral as previously described, such that every site of the square lattice is doubly occupied ($\upharpoonleft\downharpoonright$).  
(See supplement for a discussion.)
The fermions experience an on-site, inter-species repulsion $U \approx 6t$ (which may have been introduced at any stage in the protocol).
The optical lattice is now gradually morphed into a lattice with twice the number of sites~\footnote{
In Fig.~\ref{SiteDoubling} we have assumed that the bosons experience a double-well potential with appreciable intradimer tunneling, such that they occupy peanut-shaped orbitals and produce a similarly shaped Hartree potential that blocks the fermions from escaping.
}
as illustrated in Fig.~\ref{SiteDoubling}.  
This may be done by ramping up a sublattice or by ramping down a superlattice.  The feasibility of studying cold atom systems in double-well potentials is well established, having been demonstrated experimentally by several groups \cite{sebbystrabley2006,lundblad2008,trotzky2008,wirth2011,oelschlaeger2011}.
(During the preparation of this manuscript, we noted that such an idea has been proposed very recently~\cite{lubasch2011}.)

Starting from a fully occupied 2D Hubbard model at filling  $\mean{n}=2$, the site-doubling procedure leads to a 2D Hubbard model at half-filling, $\mean{n}=1$.  If this procedure is performed adiabatically, it will lead to the ground state of the final model, which is a singlet state with strong antiferromagnetic correlations.   
\footnote{
In the infinite-size limit there is a nonzero staggered moment.  In the strong-coupling limit, where the system is effectively described by a Heisenberg model, this staggered moment is $60\percent$ of its classical value.  In a finite system, however, the ground state does not break SU(2) symmetry.
}

The AF correlations will manifest themselves in the noise spectrum.
For example, snapshots of either spin species obtained using an atomic gas microscope will show a checkerboard-like pattern.
Alternatively, one may perform Bragg scattering to search for a peak at the antiferromagnetic wavevector, which is
$(\pi,\pi)$ in \emph{final} reciprocal lattice units.

\myheading{Relation to other cooling proposals:}
By the second law of thermodynamics, total entropy can never decrease.  Therefore, all cooling methods \cite{mckay2011review} operate by removing entropy from the system of interest and dumping it into a reservoir.  The reservoir may be a bath of laser photons (in the case of Doppler cooling and Zeeman slowing \cite{blochReview2008}), a gas of a different atom species (in the case of sympathetic cooling and immersion cooling \cite{ho2009squeezing}), or sacrificial atoms of the same species (in the case of evaporative cooling \cite{serwane2011}, spatial filtering \cite{popp2006,bernier2009,hoArxiv2009cooling,paiva2011,mathyArxiv2012}, band filtering \cite{blakie2007,williams2010,griessner2005}, number filtering \cite{popp2006,popp2006b}, and spin-gradient demagnetization cooling \cite{medley2011}). 
Our proposal belongs to the class of methods known as spatial filtering, in which trap and barrier potentials are manipulated to give an inhomogeneous distribution of entropy, and high-entropy regions are then isolated or discarded.  The new idea here is that instead of using lasers, one can use a second species of atoms to provide the barrier potential.  The laser approach gives optical-resolution barriers, analogous to controlling electrons using lithographically fabricated solid-state devices, whereas the atom approach gives atomic-resolution barriers, analogous to nanostructures built using a scanning tunneling microscope tip or by self-assembly.


It is sometimes unclear whether ``cooling'' refers to a reduction of temperature or of entropy.  It is worthwhile clarifying the thermodynamics of the proposed protocol as follows.
The first step (ramping up the corral potential) isolates the low-entropy gapped core from the high-entropy outer shell.  The second, optional step (discarding excess fermions) removes entropy far away from the system.  It is only during the final step (adiabatic evolution into an antiferromagnet) that the temperature of the system decreases; of course, temperature is not particularly meaningful for a small, isolated system
\cite{mckay2011review}.

\myheading{Estimates of deviations from ideal behavior:}
We finally attempt to estimate the entropy and temperature of the final state due to non-idealities at various stages of the protocol.

\begin{figure}[!htb]  \centering
		\includegraphics[width=0.6\columnwidth]{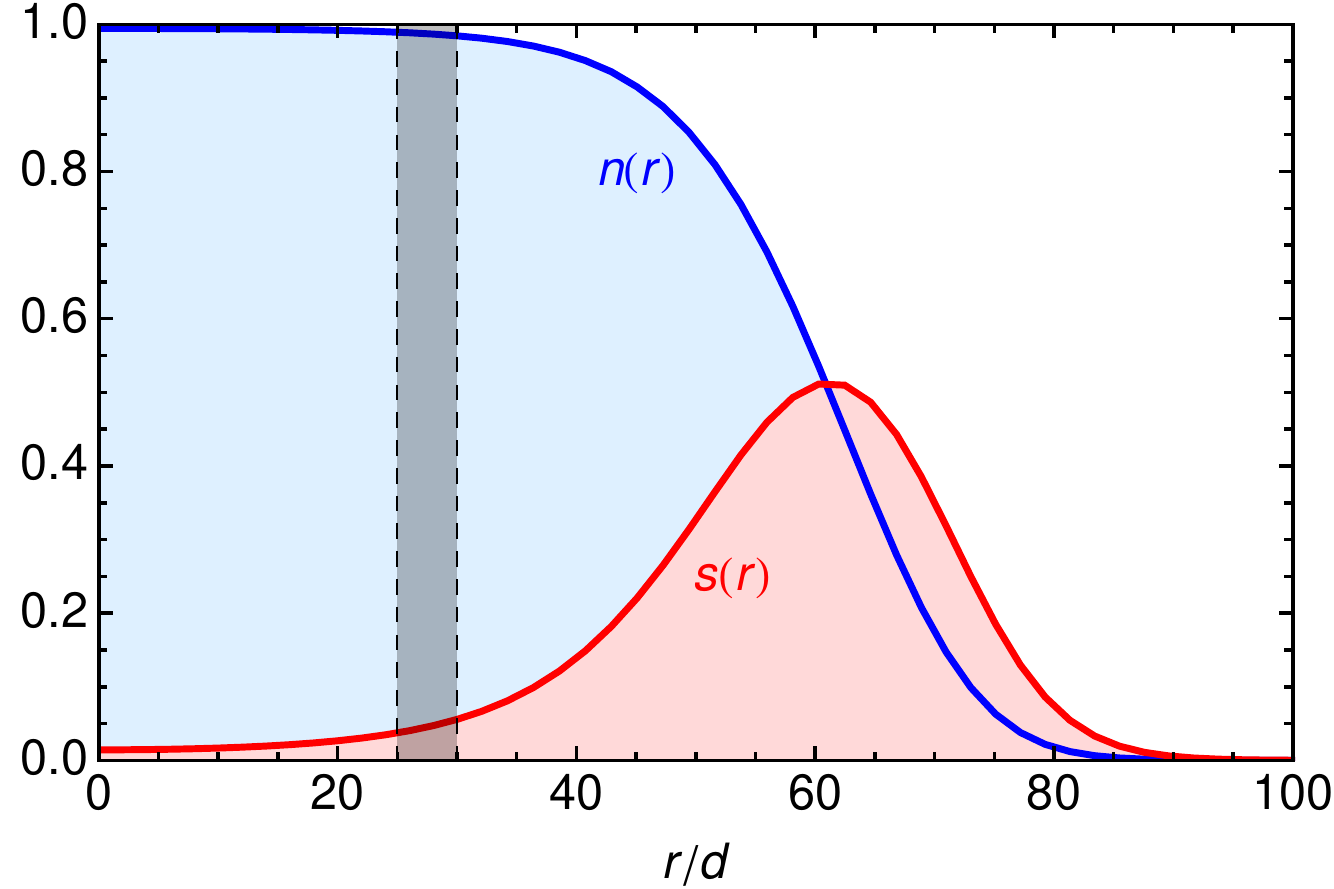}
		\caption{
			\label{VacancyDensity}
			\textbf{Density profile:}
			The blue curve is the density profile
				of non-interacting fermions 
				in a cubic lattice
				in a harmonic trap, according to the local density approximation.
			The system consists of a $n=1$ band insulator
				surrounded by a Fermi liquid outer shell with $0<n<1$
				that contains most of the entropy.
			The gray region represents a bosonic corral of inner radius $25d$
				and thickness $5d$.
		}
	\end{figure}

\myheading{(a) Imperfections in band insulator:}
The first stage of the process involves preparing a Fermi band insulator in the center of a trap (by evaporative or sympathetic cooling).
Here, we benefit naturally from entropy distribution in an inhomogeneous system:
the outer shell is a gapless Fermi liquid that can ``soak up'' entropy,
so that the core, which is a gapped Fermi band insulator, 
has a lower entropy density than average. 
As an example of typical parameters, we consider ${}^{6}$Li in a 3D cubic lattice of spacing $d=532~\mathrm{nm}$ with trap frequency $\Omega_{xy} = 2\pi \times 180~\mathrm{Hz}$ and lattice depth $V_0 = 8 E_R$.  In this case the nearest-neighbor hopping amplitude is $t_\text{hop}/k_B \approx 43~\mathrm{nK}$ and the trap constant is $\alpha \approx 0.003t_\text{hop}$, such that $V(r) = \alpha (r/d)^2$. 
For a non-interacting Fermi gas with $N=10^6$ particles 
and overall entropy per particle $S/N=0.65k_B$, 
the density and entropy profiles are shown in Fig.~\ref{VacancyDensity}.
Using a spherical corral of radius $25d$ we can isolate a core subsystem with vacancy concentration $0.008$ and average entropy per particle $0.03k_B$.  
While this entropy is not strictly zero, it does represent a 25-fold reduction from the starting average entropy, and it is certainly well below the critical entropy for antiferromagnetic ordering in 3D~\cite{fuchs2011,paiva2011}.  In fact, the initial entropy does not impose a fundamental theoretical limit on the entropy in the corral~\footnote{The entropy density decays exponentially in the insulating core.  The extent of this core is limited by the gap to the second Bloch band, but this gap can be made arbitrarily large by using a deep lattice.  The depth of the lattice is limited by the need for the fermions to hop on the timescale of the experiment; but this is a practical limitation rather than a fundamental one.
}.

For a 2D square lattice the entropy redistribution effect is less dramatic (because the size of the Fermi liquid region is amplified by the Jacobian, which is now only $2\pi R$ instead of $4\pi R^2$).  On the other hand, it should be easier to reach lower initial temperatures for a 2D system by evaporative cooling, because one can evaporate away all the atoms that are not in the plane of interest.

	\begin{figure}[h]  \centering
		\includegraphics[width=50mm]{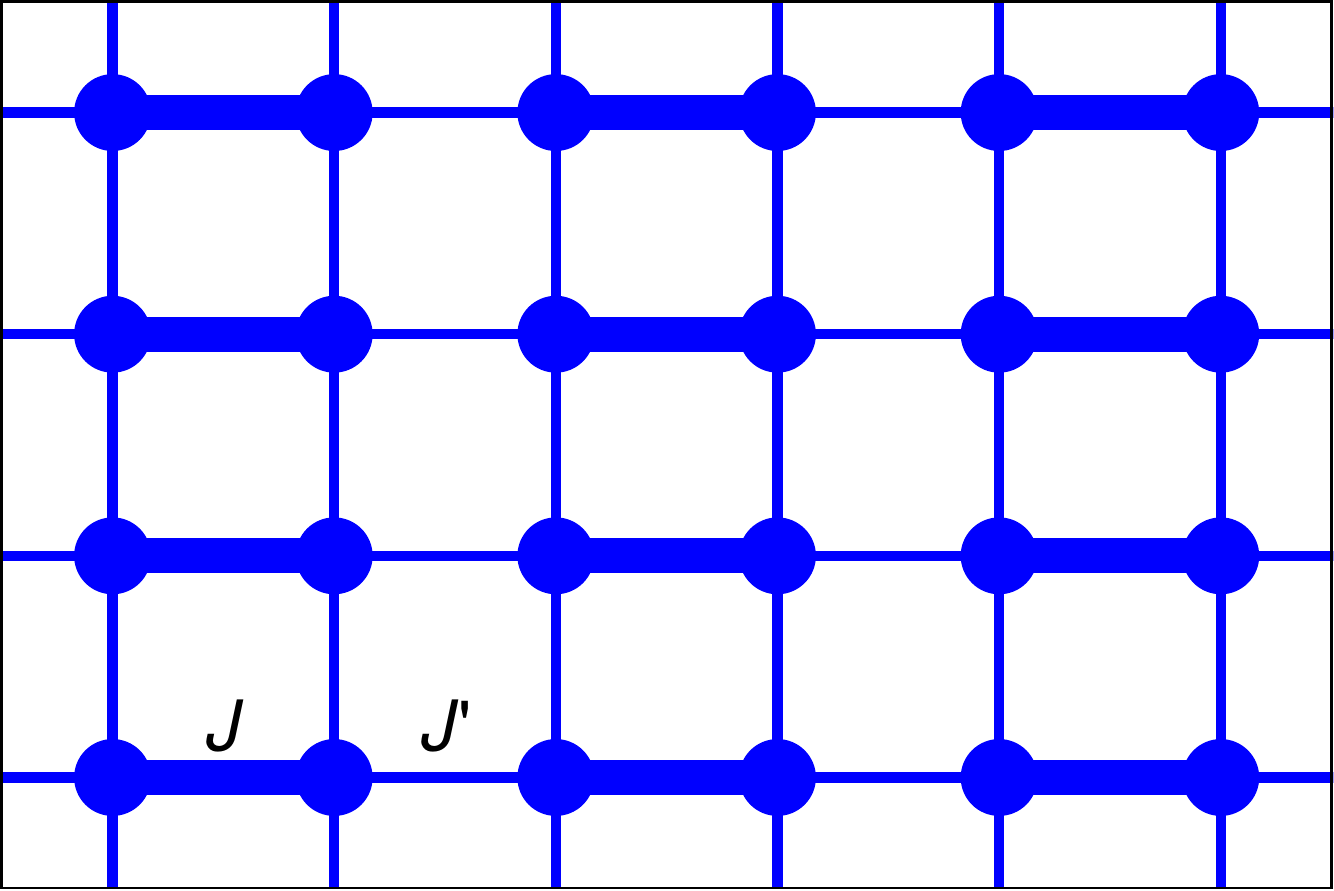}
		\caption{
			\label{MyHeisenbergModel}
			\textbf{Heisenberg model on dimer lattice:}
			Thick bonds represent exchange coupling $J$ 
			and thin bonds represent coupling $J' < J$.
		}
	\end{figure}

	\begin{figure*}  \centering
		\includegraphics[width=\textwidth]{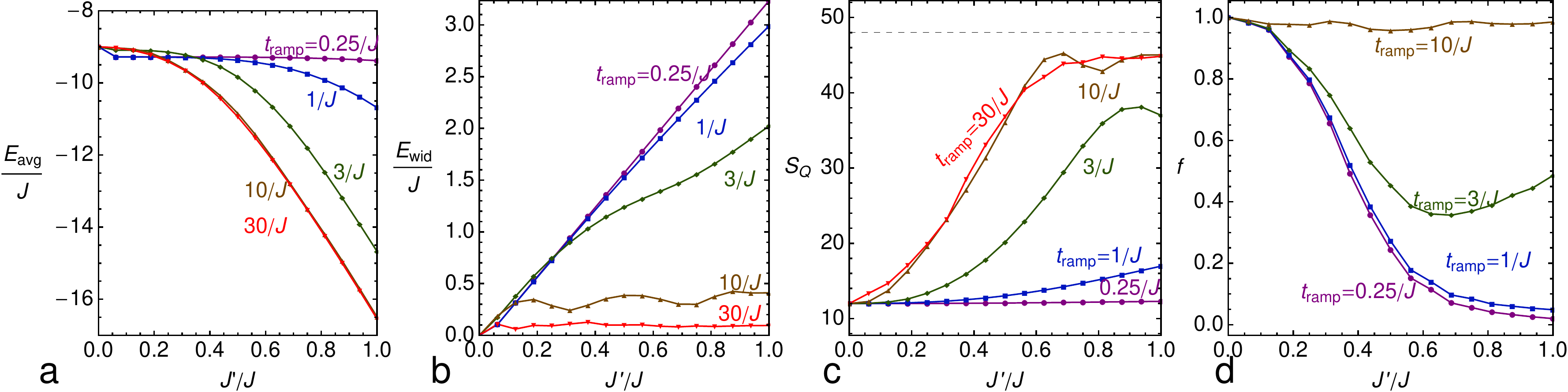}
		\caption{
			\label{TimeEvolution}
			\textbf{Time evolution of various quantities:}
			Mean energy $E_\text{avg}$,
				energy uncertainty $E_\text{wid}$,
				staggered structure factor $S_\QQQ$,
				and fidelity $f$
				on a $6\times 4$ lattice
				as a function of interdimer coupling $J'$ during the AF-generating protocol
				for various ramp times $t_\text{ramp}$.
			For $t_\text{ramp} \lesssim 1/J$ (sudden turn-on),
				the wavefunction has no chance to evolve,
				so the AF structure factor $S_\QQQ$ stays constant
				and the energy uncertainty grows quickly.
			For $t_\text{ramp} \gtrsim 10/J$ (adiabatic ramp-up),
				$S_\QQQ$ grows indicating the growth of AF correlations
				across the quantum critical point,
				$E_\text{avg}$ tracks the ground state energy,
				and $E_\text{wid}$ remains small, 
				indicating high fidelity (negligible admixture of excited states).
			The black dashed line in (c) indicates the structure factor for 
				a classical N\'eel antiferromagnet $S_\QQQ=N_\text{sites}^2/12$.
			The unit of energy is the intra-dimer coupling $J$.
			For a comparison between different lattice sizes,
				see supplement.
		}
	\end{figure*}

\myheading{(b) Non-adiabaticity in site-doubling process:}
We now study the possibility of adiabaticity/non-adiabaticity in the process shown in Fig.~\ref{SiteDoubling}.
As we go from isolated dimers (Fig.~\ref{SiteDoubling}b) 
towards a rectangular lattice (Fig.~\ref{SiteDoubling}c),
the system remains half-filled with one fermion per site on average,
and the ``charge gap'' remains robust.
Thus, for the purpose of modeling the process we focus on the spin degrees of freedom.
We consider a SU(2) $S=1/2$ Heisenberg antiferromagnet on a lattice of dimers, with coupling $J$ within each dimer and $J'$ between dimers,
	\begin{eqnarray}
	H_\text{Heis} 
	= 
		  \sum_{ij \in \calL}  J \SSS_i \cdot \SSS_j
		+ \sum_{ij \in \calL'}  J' \SSS_i \cdot \SSS_j
	,
	\label{MyHeisenbergModelEqn}
	\end{eqnarray}
where $\calL$ and $\calL'$ are the two sets of bonds shown in Fig.~\ref{MyHeisenbergModel}
\footnote{
There have been quantum Monte Carlo studies of the equilibrium phase diagrams of similar models (such as coupled two-leg ladders).
Here we perform a simple calculation on the dynamics of small systems to obtain an idea of the timescales required for adiabaticity.
}.
We start with $J=1$ and $J'=0$ (isolated dimers).  The ground state is a spin-gapped singlet state with singlet ``valence bonds'' living on strong bonds $J$, which we shall call a ``valence bond solid'' (VBS) for want of a better term,
	\begin{eqnarray}
	\ket{\psi(t=0)} 
	&= 
		\bigotimes_{ij \in \calL_1}
		\tfrac{1}{\sqrt{2}} \big(
			\ket{\up}_i \ket{\dn}_j - \ket{\dn}_i \ket{\up}_j 
		\big)
		.
	\nonumber
	\end{eqnarray}
We represent it in the $S_Z$ eigenbasis, as a vector with $2^{N_\text{sites}}$ coefficients.
The system is now ramped from isolated dimers ($J'=0$) to an isotropic square lattice ($J'=1$) as a function of time $t$ over a total time $t_\text{ramp}$.
We evolve the wavefunction by solving the time-dependent Schr\"odinger equation 
	$
	\frac{\dd}{\dd t} \ket{\psi(t)} 
	= 
		-i \hat{H} (t) \ket{\psi(t)} 
	$
with the Verlet method,
where $\hat{H}$ and $\ket{\psi}$ are time-dependent
 (see supplementary information).
As the lattice becomes more isotropic, antiferromagnetic correlations grow,
which is an indication of the quantum phase transition to an antiferromagnetic state in a bulk system.

The staggered structure factor 
$S_\QQQ =	\langle 	\hat{M}_\QQQ {}^2	 \rangle $, where 
$\hat{M}_\QQQ =  \sum_{i} (\exp i\QQQ\cdot\RRR_i) \hat{S}^Z_i$
and $\QQQ \equiv (\pi,\pi)$, 
indicates the size of antiferromagnetic fluctuations.
With this definition the ``VBS'' state has $S_\QQQ=N_\text{sites}/2$,
whereas a classical N\'eel antiferromagnet with a N\'eel vector $\MMM_\QQQ$ in the $z$ direction would have $S_\QQQ=N_\text{sites}{}^2/4$.  Averaging over all directions of $\MMM_\QQQ$ on the surface of a sphere reduces this to $S_\QQQ=N_\text{sites}{}^2/12$.  Quantum fluctuations further affect this number.  

We also calculate the mean energy 
$E_\text{avg} = 	\bra{\psi}  \hat{H}  \ket{\psi}$
and the energy variance 
$E_\text{wid} {}^2	=	\bra{\psi}  \hat{H}^2  \ket{\psi} 	-	\bra{\psi}  \hat{H} \ket{\psi}^2$.
$E_\text{wid}$ is a measure of the admixture of excited states produced by non-adiabaticity; roughly speaking, it is also an indication of the eventual temperature due to heating due to non-adiabaticity (although we do not model the thermalization process here).

Figure~\ref{TimeEvolution} shows $E_\text{avg}$, $E_\text{wid}$, and $S_\QQQ$
as a function of time $t$ 
for various ramp rates $1/t_\text{ramp}$.
For sufficiently slow ramp times the behavior of these quantities shows that the evolution is adiabatic (see caption).
This allows us to estimate the fidelity (the squared overlap of the wavefunction with the instantaneous ground state) as
$f(t_\text{ramp}, J') 
= \left| \braket{ \psi(30, J') }{ \psi(t_\text{ramp}, J') } \right|^2$,
shown in the last panel of Fig.~\ref{TimeEvolution},
which gives further evidence for our results.

For a hopping amplitude $t_\text{hop} = 2~\mathrm{nK}$ and a Hubbard repulsion $U = 8t_\text{hop}$, the exchange coupling scale is $J \sim 4t_\text{hop}^2/U = 1~\mathrm{nK}$, corresponding to a time scale of $1/21~\mathrm{Hz} = 50~\mathrm{ms}$.  Thus, the ramp time should be at least $500~\mathrm{ms}$.
This is short enough to permit measurements in the antiferromagnetic state to be made before the state is destroyed by atom losses and/or technical heating.

Adiabaticity is relatively easy to achieve in the disordered phase, where there is a spin gap.
In a bulk system the spin gap decreases to zero at the QCP, and the excitation density scales as a power law of the ramp time, governed by appropriate critical exponents~\cite{polkovnikov2005,zurek2005,dziarmaga2005}.
However, in a finite system of linear size $L$, the gap at the QCP scales as $L^{-1}$ (because the dynamic critical exponent is $z=1$).
Therefore the ramp time necessary for adiabaticity scales only as $L$.
This is a likely explanation for the observations of Ref.~\onlinecite{lubasch2011}).
For sufficiently large ramp times the excitation density is exponentially small 
according to Landau-Zener theory.
This is an encouraging observation.

In conclusion, we have proposed a very promising method for obtaining quantum ground states.
The experimental challenge is to find a system where the lattices and interactions can be tuned appropriately. 

YLL acknowledges support from DARPA grant no. 60025344 under the Optical Lattice Emulator (OLE) program and is grateful to William Cole, Eric Duchon, Richard Scalettar, and Marc Cheneau for helpful discussions.




\end{document}


\title{Cooling by corralling: a route to ultra-low entropies in optical lattices\\
Supplementary Information}
\maketitle

\subsection{Removal of excess fermions: details}
Na\"ively, one might think that momentary application of a steep inverse harmonic potential would quickly cause the exterior fermions to fly away in all directions.  However, when the fermions have accelerated to the Brillouin zone edge (with crystal momentum $k=\pi/d$) they Bragg-scatter to the opposite edge ($-\pi/d$), the group velocity changes sign, and the fermions start moving \emph{inwards} instead (this is the phenomenon of Bloch oscillations).

A more suitable approach is to apply the anti-trapping potential for a short duration to impart an outward crystal momentum to the exterior fermions
(noting that the maximum group velocity occurs in the middle of the dispersion relation),
and then turn off the potential to allow the fermion cloud to expand ballistically.

\begin{figure*}
		\includegraphics[width=\textwidth]{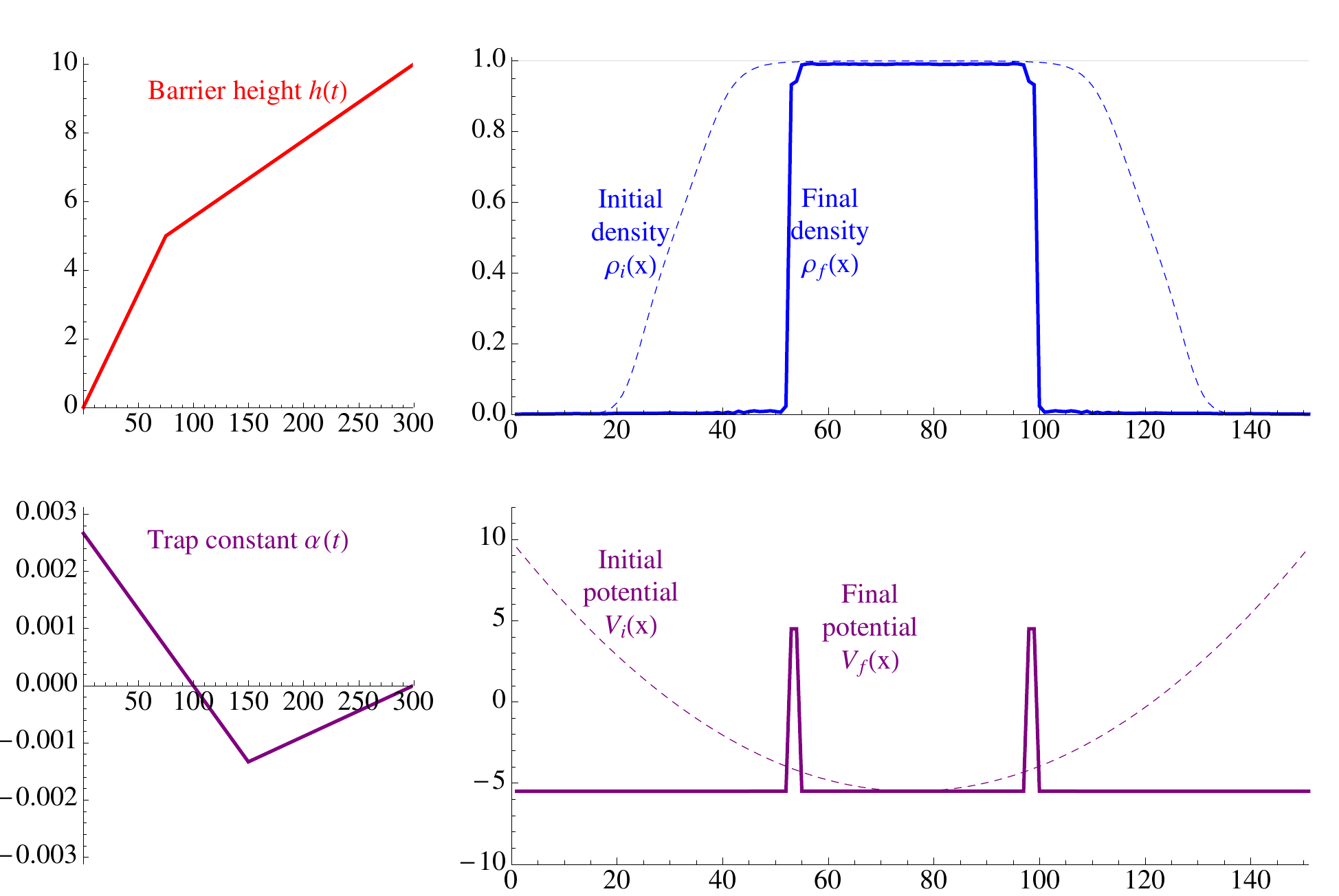}
		\caption{
			\label{TrapEmptyingSimulation2}
			Potential $V(x)$ (top panel) and density profile $\rho(x)$ (bottom)
				for 90 fermions in a 151-site tight-binding model
				with a time-dependent potential.
			Dashed curves show the potential and ground state density at time $t=0$.
			Solid curves show the potential and density at $t=40$,
				when the excess fermions have moved far away from the corral.
		}
	\end{figure*}

To confirm the feasibility of this idea it suffices to study the problem in 1D.
Figure~\ref{TrapEmptyingSimulation2} shows a simulation of the dynamics of 90 non-interacting spinless fermions in a 151-site tight-binding chain.
The unit of energy is the nearest-neighbor hopping $t_\text{hop}=1$, and the unit of length is the lattice spacing $d$.
(In a typical experiment $\hbar/t_\text{hop} \sim 25~\mathrm{ms}$ and $d \sim 600~\mathrm{nm}$.)

\myheading{Simulation of quantum dynamics of many non-interacting fermions in a time-dependent potential:} 
We initialize the system in a harmonic potential $V(r)=\alpha(0) r^2$ where $\alpha(0)=0.00267$ and $r$ is the distance from the center of the trap in lattice spacings.  We assume that the system starts off in thermal equilibrium at temperature $T=0.5$ and chemical potential $\mu=5.5$.  We construct the density matrix, $\hat{\rho} = f((\hat{H}-\mu)/T)$, where $f$ is the Fermi function.
(Here we have taken the density matrix formalism, which is usually applied to an ensemble of independent systems, and co-opted it for non-interacting fermions living together in one system -- hence the Fermi factor instead of a Boltzmann factor.)
Then we gradually ramp up two square potential barriers with height $h(t)$ while simultaneously relaxing the harmonic trap constant according to $\alpha(t)$.  
In real experiments the optical lattice beam has a finite waist.  Therefore, once the atoms have travelled a sufficient distance from the trap center, they no longer experience Bloch oscillations and they should quickly be lost from the system.  In order to model this without too much effort, we add a small imaginary part $i\Gamma(x)$ to the potential at four end sites.
We simulate the time-dependent Schr\"odinger equation by applying a succession of unitary time-evolution operators $e^{-i \hat{H}(t) \delta t}$ as similarity transformations to the density matrix $\hat{\rho}$, where the timestep is $\delta t=0.3$, from time $t=0$ until $t=300$. 
See ancillary file (\texttt{TrapEmptyingAnimation2.mov}) for an animation of the evolution of the potential and density profile.

The final state has approximately 47 fermions.  The interior of the corral is almost a perfect $n=1$ band insulator in a flat-bottomed potential well; there is negligible entropy accumulated (or generated) within the corral.  The barrier sites themselves have fractional occupation (and hence finite entropy), but the large potential difference between barrier sites and interior sites completely prevents this entropy from leaking into the corral.  

In this simulation the parameters were fairly conservative -- the corral was chosen to be small in order to capture little entropy.  One could attempt using a larger corral to capture more fermions, at the cost of also capturing more entropy.

\myheading{Considerations for choosing a barrier ramp rate:} 
It is impossible to ramp up the barrier(s) adiabatically.  According to Landau-Zener theory, adiabaticity requires a ramp slower than the timescale of the smallest excitation energy -- the splitting corresponding to the tunneling amplitude through the barrier -- which is extremely small once the barrier becomes moderately high.  Another way to see this is to consider an initial density close to $n=1$ in a large region on both sides of the barrier.  If the barrier is ramped up beyond the hopping bandwidth, adiabaticity requires the barrier sites to be vacated; however, most barrier fermions are likely to remain stuck on barrier sites due to Pauli blocking from all around.  

This form of non-adiabaticity is actually desirable: if all the barrier fermions remain in place, there is no entropy generated even at the barrier!

If the barrier is raised faster than the timescale corresponding to the gap between the first and second Bloch bands, fermions may be promoted into higher bands.  Then the simulations presented here, which assume a single-band tight-binding model, are no longer valid.  We have not examined this situation in detail.


There are other possible ways of removing excess fermions, such as using appropriate radiofrequency or laser pulses (or the orbital excitation blockade) to ``blast away'' fermions at corral sites.

\subsection{Evolution into antiferromagnetic state: details}
\begin{figure}[!h]
		\includegraphics[width=\columnwidth]{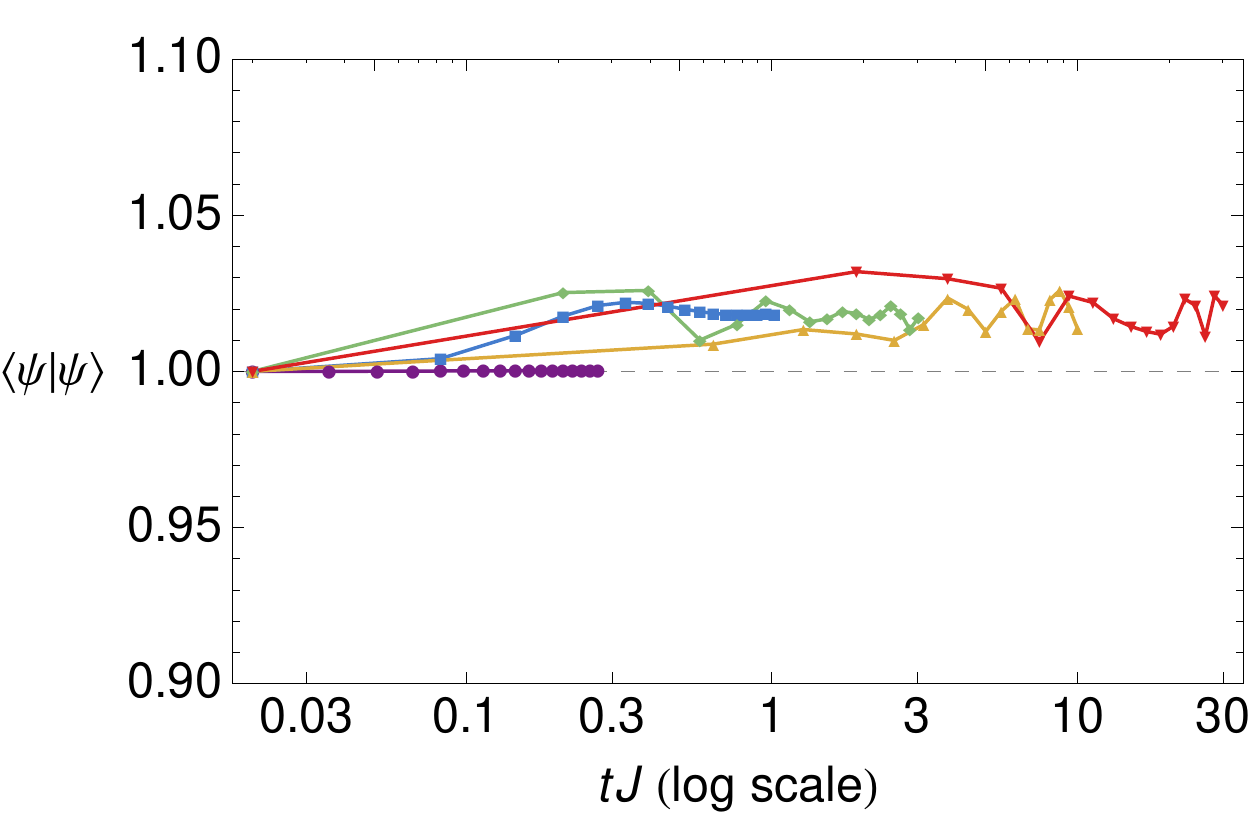}
		\caption{
			\label{TimeEvolutionConservesWavefunctionNorm}
			Time evolution of the norm of the wavefunction on a 6x4 lattice
			for various ramp times $t_\text{ramp} = 0.25/J$ to $30/J$.
		}
	\end{figure}

We used the Verlet method for time evolution, in which the wavefunction is split into real and imaginary parts, $\boldsymbol{\psi} = \uuu + i\vvv$, and evolved according to a second-order discretization \emph{of the Schr\"odinger equation},
	\begin{align}
	\uuu(t+\half \delta t) 	&= \uuu(t) - i \HHH(t+\half \delta t) \vvv(t)   \nonumber\\
	\vvv(t+\delta t) 	&= \vvv(t) + i \HHH(t+\half \delta t) \uuu(t+\half \delta t)   \nonumber\\
	\uuu(t+\delta t) 	&= \uuu(t+\half \delta t) - i \HHH(t+\half \delta t) \vvv(t+\delta t)
	.
	\end{align}
The timestep $\delta t$ was made small enough so that the oscillations in the norm of the wavefunction, $\braket{\psi}{\psi}$, were smaller than $5\percent$, as shown in Fig.~\ref{TimeEvolutionConservesWavefunctionNorm}.

An alternative method involves discretizing the \emph{time-evolution operator} as $\prod_t  e^{-i\VVV(t) \delta t/2} e^{-i\KKK(t) \delta t} e^{-i\VVV(t) \delta t/2}$.  
This discretization conserves the norm of the wavefunction exactly, but there is still an unknown $O(\delta t^2)$ ``phase error.''  Ironically, by eliminating amplitude errors, one loses the ability to estimate phase errors (which is crucial to check the validity of the results)!  Thus in this paper we stick to the Verlet method.

\begin{figure*}[p]
		\includegraphics[width=\textwidth]{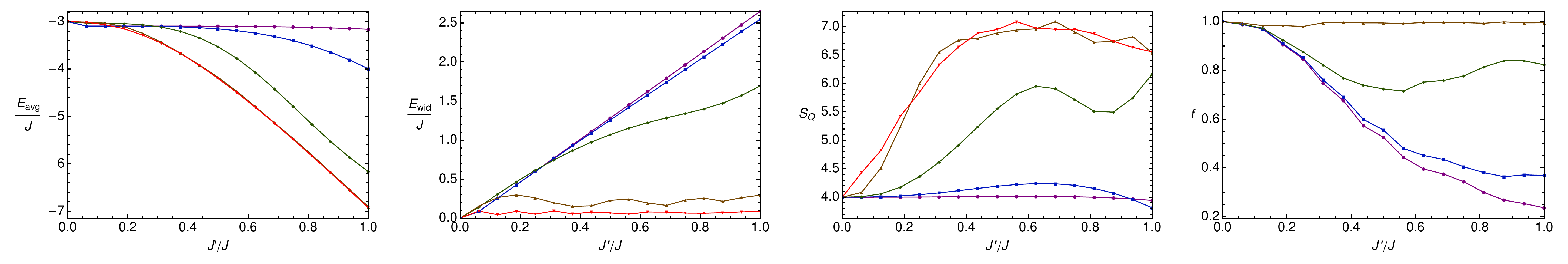}
		\includegraphics[width=\textwidth]{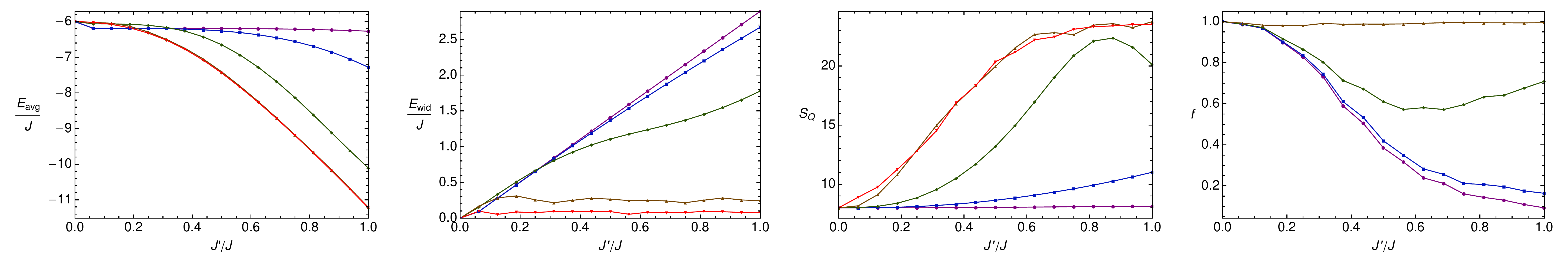}
		\includegraphics[width=\textwidth]{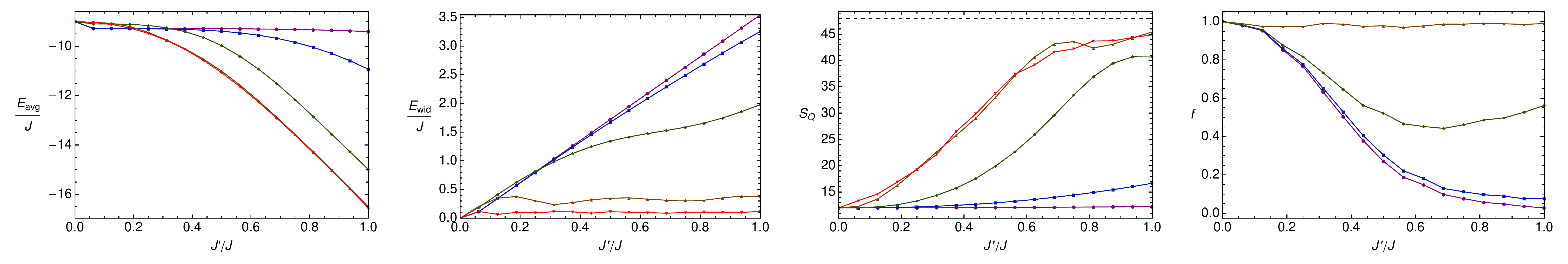}
		\includegraphics[width=\textwidth]{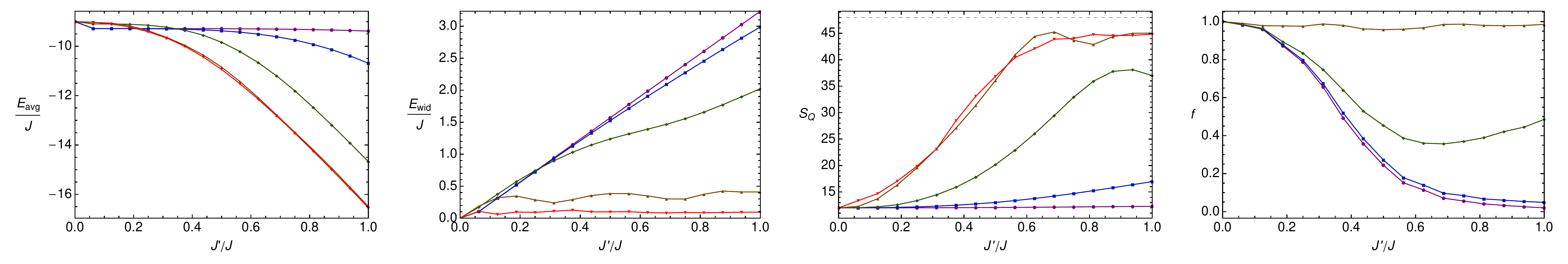}
		\caption{
			\label{TimeEvolutionSystemSizeDependence}
			Time evolution of
				mean energy $E_\text{avg}$, 
				and energy uncertainty $E_\text{wid}$,
				staggered structure factor $S_\QQQ$, 
				and fidelity $f$
				as a function of time-dependent coupling $J'$,
				for various ramp times $t_\text{ramp} J=0.25,1,3,10,30$ (colors),
				on lattices of various sizes.
			The lattice sizes, from top to bottom, are
				$4\times 2$, $4\times 4$, $4\times 6$,
				and $6\times 4$ (the last of which was presented in the main text).
			Dashed lines on $S_\QQQ$ plots indicate $N_\text{sites}^2/12$,
				the value expected for a classical N\'eel antiferromagnet 
				with staggered moment averaged over a sphere.
			On small lattices the true $S_\QQQ$ can exceed this classical value
				due to the peculiar nature of quantum mechanical spin.
		}
	\end{figure*}

Figure~\ref{TimeEvolutionSystemSizeDependence} compares the time evolution of the observables $E_\text{avg}$, $E_\text{wid}$, $S_\QQQ$, and $f$ for various system sizes.  
We see that for a ramp time $t_\text{ramp} = 10/J$ the energy uncertainty $E_\text{wid}$ is very close to zero and the fidelity $f$ is very close to unity, indicating adiabaticity.

\subsection{Other methods for adiabatic evolution to an AF}
In order to obtain interesting states of fermions one has to have at least two species of fermions.  The simplest version of our proposal produces only one species of fermions.
The easiest way to generate two species of fermions is to apply a $\pi/2$ pulse to evolve the fermions from the original hyperfine state $\ket{1}$ to into a superposition $\frac{1}{\sqrt{2}} ( \ket{1} + \ket{2} )$, and then ramp down the fermionic lattice depth ($V_f$) to allow tunneling between sites.  However, this approach involves a Hamiltonian with a massively degenerate ground state (now that we are in the Fock subspace with $N_1 = N_2 = N_\text{sites} /2$), from which adiabatic evolution is useless.
If instead we begin with fermions in a shallow lattice permitting intersite tunneling, then the coherent superposition $\frac{1}{\sqrt{2}} ( \ket{1} + \ket{2} )$ is \emph{not} the ground state of the Hamiltonian, but rather, it is a fully polarized state (polarized along the $x$-axis in pseudospin space), which will eventually relax to an unpolarized two-component Fermi liquid, turning kinetic energy into heat in the process.  
Due to the above difficulties, we propose using a two-component band insulator as a starting state and performing a site-doubling procedure as described in the text.

	\begin{figure}[!htb]
		\includegraphics[width=\columnwidth]{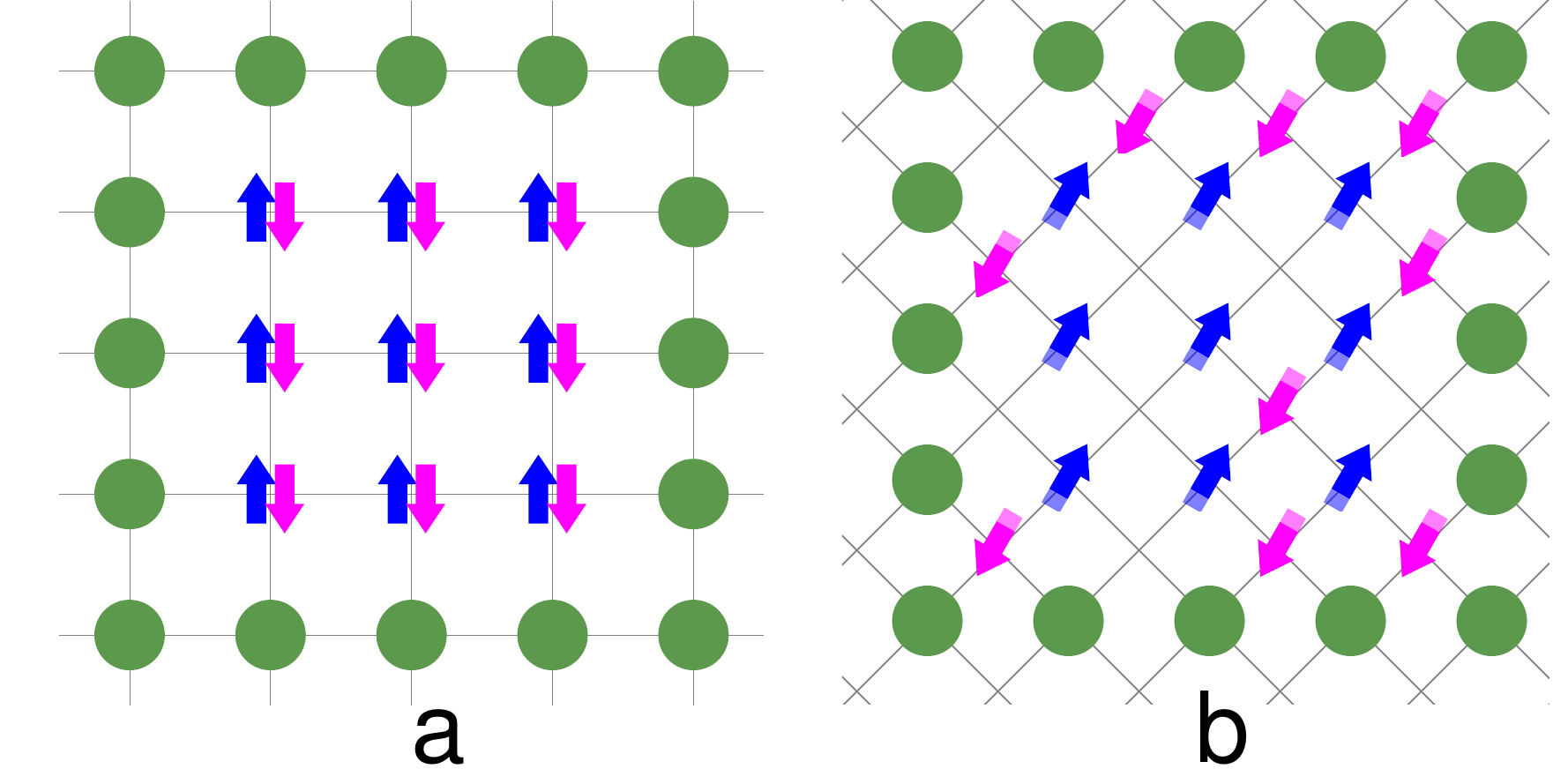}
		\caption{
			\label{SiteDoublingAlternative}
			An alternative site-doubling procedure,
			in which a sublattice is gradually ramped up
			to introduce sites in the centers of the original unit cell.
		}
	\end{figure}

In the text we considered site doubling using a double-well geometry.
A possible alternative involves morphing from a square lattice with a $d \times d$ unit cell to a $45^\circ$ $\frac{d}{\sqrt{2}} \times \frac{d}{\sqrt{2}}$ lattice,
as illustrated in Fig.~\ref{SiteDoublingAlternative}.
(Here we assume that the new lattice only applies to the fermions.)
In this case, the number of sites in the new lattice is not exactly twice the original number of sites due to edge effects, so the final state is a hole-doped antiferromagnet.
On the other hand, this geometry might be less susceptible to leakage (fermions slipping past bosons).  The hole-doped antiferromagnet itself is, of course, an extremely important and interesting system to study using cold atoms.

\subsection{Other methods for analyzing non-adiabaticity}
Apart from analytic theories (Landau-Zener and Kibble-Zurek), one can study the dynamics of large systems using time-evolving block decimation (TEBD) methods that use a matrix product state representation of the wavefunction.  However, this approach is limited to 1D, so we have not explored it further.

\subsection{Generalizations}
There is a great deal of flexibility in the implementation of the corral idea.  The corral can be formed either from a Bose Mott insulator, as described in the text, or from a Fermi insulator with one fermion localized at every site.  Similarly, the corralled species can be either a Bose MI or a Fermi BI (as described).

As state in the text, the idea applies in 1D, 2D, and 3D.  It is interesting to note that for a 1D chain, two groups of bosons (e.g., at sites $x=1,2,3,4,21,22,23,24$) are sufficient to form a corral.  This situation is particularly easy to set up.  However, antiferromagnetic order is impossible in 1D.




